\documentclass[prl,10pt,twocolumn,superscriptaddress,aps,amsmath,amsfonts,notitlepage,longbibliography]{revtex4-2}
\usepackage{amsmath,amsfonts,amssymb,dsfont,graphicx,bm}
\graphicspath{{Figures/}}
\usepackage{color}
\usepackage{tikz}
\usepackage{pbox}
\usepackage{balance}
\usepackage{hyperref}

\usepackage{txfonts}

\usepackage{xr}
\usepackage[normalem]{ulem}

\def\be{\begin{equation}}
\def\ee{\end{equation}}
\def\bea{\begin{eqnarray}}
\def\eea{\end{eqnarray}}
\def\bpm{\begin{pmatrix}}
\def\epm{\end{pmatrix}}

\def\Im{\mathop{\rm Im}}
\def\Re{\mathop{\rm Re}}

\def\diff{\mathrm{d}}
\def\Diff{\mathrm{D}}

\newcommand{\p}{\partial}

\begin{document}
\title{Vestigial altermagnetism}

\author{Peng Rao}
\affiliation{Technical University of Munich, TUM School of Natural Sciences, Physics Department, 85748 Garching, Germany}
\author{Johannes Knolle}
\affiliation{Technical University of Munich, TUM School of Natural Sciences, Physics Department, 85748 Garching, Germany}
\affiliation{Munich Center for Quantum Science and Technology (MCQST), Schellingstr. 4, 80799 München, Germany}
\date{\today}
\begin{abstract}

Interaction effects and fluctuations govern the phase diagram of systems with multiple competing phases. In particular, as individual phases are destroyed by increasing temperature, fluctuations can still produce a composite `vestigial' order given by a product of individual order parameters. Here, we describe vestigial Altermagnetism (AM) from the thermal melting of antiferromagnetic (AFM) and orbital orders (OO) on the square lattice. The coexistence of AFM and OO can give rise to conventional AM with ordering wave vector $\mathbf{Q}=(\pi,\pi)$. We show that a distinct `vestigial AM' phase with $\mathbf{Q}=0$ can arise from thermal melting of AFM and OO, and investigate its generic phase diagram. We find that the transition from the high-temperature disordered phase into the vestigial AM phase can be either first or second order, depending on the AFM and OO bare susceptibilities and their interactions. We then investigate the unique experimental signatures by calculating the electron spectral function for a minimal two-orbital model. Despite the fact that the vestigial AM order parameter does not couple to electrons directly, fluctuations dominate the self-energy corrections giving rise to a spin- and orbital-dependent quasi-particle lifetime with the AM symmetry.


\end{abstract}

\maketitle

\textit{Introduction.} 
Fluctuation phenomena are important in systems hosting multiple competing phases, where the interplay between order parameters can produce new and exotic phases of matter~\cite{fradkin2015colloquim}. In particular, above the ordering temperatures of individual phases, their order parameters vanish but fluctuations can act as a glue forming a composite phase described by a product of individual order parameters. Such `vestigial' orders are a precursor of the individual ordered phases with lower symmetry~\cite{fernandes2019intertwined}. A celebrated example in the context of frustrated magnetism is the square lattice $J_1$-$J_2$ Heisenberg model with its low temperature stripe phase breaking both spin and rotational lattice symmetry. Its vestigial phase at higher temperature only breaks the lattice symmetry forming a 'lattice nematic' described by the Ising variable $\langle \mathbf{S}_1.\mathbf{S}_2\rangle$, where $\mathbf{S}_1$ and $\mathbf{S}_2$ are spins from the two sublattices coupled by the nearest-neighbour $J_1$~\cite{chandra1990ising}. Over the last years the concept of vestigial order has also been employed to describe unconventional superconductivity~\cite{agterberg2008dislocations,wang2015coexistence,cai2017interwined}, in particular charge $4e$-superconductivity as a composite order $\langle \Delta^2 \rangle $ of the gap functions $\Delta$~\cite{berg2009charge}, as well as exotic charge, spin and nematic orders in the Iron-based compounds~\cite{eremin2010magnetic,brydon2011microscopic,giovannetti2011proximity,fernandes2012preemptive,fernandes2014drives,wang2015magnetic,fernandes2016vestigial}.


 \begin{figure}[t!]
    \centering
    \includegraphics[width=0.85\linewidth]{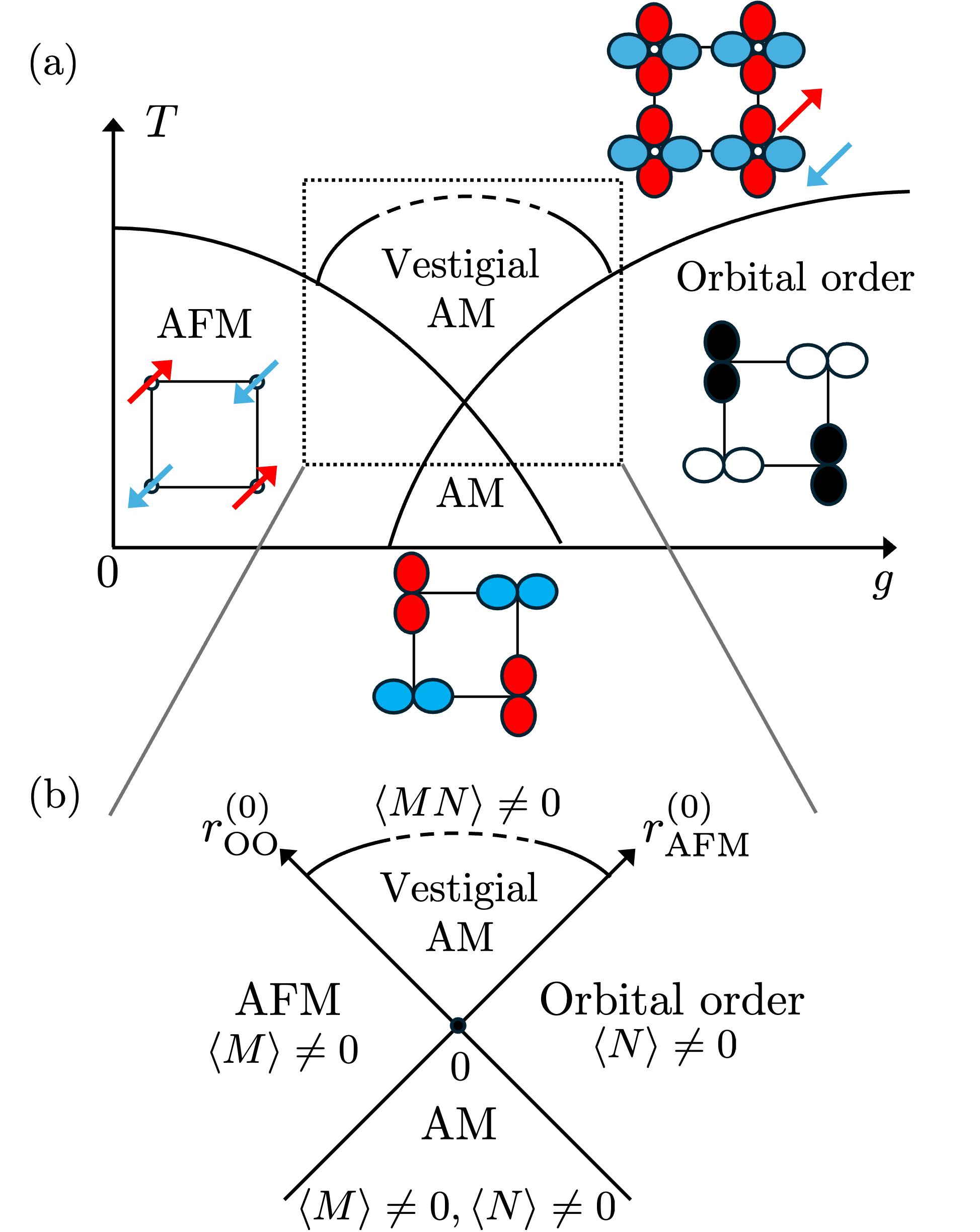}
    \caption{(a) The phenomenological phase diagram as a function of temperature $T$ and tuning parameter $g$. The $x$- and $y$-orbitals are shown as horizontal and vertical dumbbells. Red dumbbell means the electrons occupying the orbital are spin-up, and blue dumbbells are for spin-down electrons. Above the long-range ordered coexistence phase of the AM a distinct vestigial AM appears from thermal fluctuations. Along the dashed curve, the transition to the vestigial AM phase is first order. (b) the zoomed-in mean field phase diagram [gray box in (a)] as a function of the susceptibilities $r_{\text{AFM}}^{(0)}, r_{\text{OO}}^{(0)}$. The AFM and OO is present at $r_{\text{AFM}}^{(0)}<0$ and $r_{\text{OO}}^{(0)}<0$ respectively.}
    \label{fig:phasediagram_schematic}
\end{figure}

Thermal melting of competing phases and vestigial order can arise naturally in altermagnetism (AM), which is proposed as a third type of collinear magnetism distinct from ferromagnetism and antiferromagnetism~\cite{smejkal2022am,smejkal2022am1}. Altermagnets have antiferromagnetic (AFM) order but contrary to antiferromagnets, are invariant under combined lattice rotation and time-reversal transformations only. Typically, the combined symmetry is attributed to the inequivalent crystal environments for the magnetic atoms on different sublattices. However, it has been shown that AM can also arise spontaneously due to electron interactions and the coexistence of AFM and orbital ordering (OO)~\cite{leeb2024spontaneous,kaushal2026spontaneous}. A minimal example is given by the square lattice two-orbital model, where the electron operator $\Psi_{\lambda s}$ has spin $s$ and orbital $\lambda=x,y$ degrees of freedom. Both AFM and OO with ordering wave vector $\mathbf{Q} = (\pi,\pi)$ can appear simultaneously~\cite{leeb2024spontaneous}:
\begin{equation}
    \langle M\rangle  =  \langle \Psi^\dagger s^z \Psi\rangle, \ \langle N\rangle = \langle\Psi^\dagger \tau^z \Psi \rangle ,
\end{equation}
where $s^z$ and $\tau^z$ are Pauli matrices in spin and orbital space. The electrons then occupy predominantly the spin-up $x$-orbitals on one sublattice and spin-down $y$-orbitals at the other; see Fig.~\ref{fig:phasediagram_schematic}(a). Both C$_4$ rotation and time-reversal T symmetries are spontaneously broken and the system is only invariant under the combined C$_4$T transformation, consequently, forming an AM with ordering wave vector $\mathbf{Q}$. Beyond the minimal two-orbital model, Ref.~\cite{kaushal2026spontaneous} considered a three-orbital model with more realistic microscopic interactions where the AM is stabilized by an additional $z$-orbital. In both cases, the schematic and generic phase diagram is given in Fig.~\ref{fig:phasediagram_schematic}(a) where $T$ is the temperature and $g$ is a microscopic tuning parameter of the system. For example, in Ref.~\cite{kaushal2026spontaneous} $g$ is the interorbital ferromagnetic Hund's coupling; compare with Fig.~4(c) there. Our key insight is that for increasing temperature the system does not simply melt to a trivial paramagnet once individual AFM and OO parameters vanish. Rather, the thermal melting can give rise to a vestigial composite order given by $\varphi \propto \langle M N\rangle$. Crucially, a nonzero $\varphi$ breaks C$_4$ and T symmetries and is invariant under combined C$_4$T only, i.e. the system has altermagnetic symmetries.

Note, the vestigial AM is qualitatively different from the foregoing conventional AM since its ordering wave vector is zero, as $M,N$ each carries momentum $\mathbf{Q}$, and $2\mathbf{Q}=0$ on a square lattice. It corresponds to a different AM phase with spin-up electrons occupying $x$-orbitals and spin-down electrons occupying $y$-orbitals in equal numbers at each site; see Fig.~\ref{fig:phasediagram_schematic}(a).  Ref.~\cite{giuli2025altermagnetism}  previously proposed such an AM with the bilinear order parameter $\eta \propto \langle \Psi^\dagger \tau^z s^z\Psi \rangle =\langle \Psi_x^\dagger s^z\Psi_x \rangle -\langle \Psi_y^\dagger s^z\Psi_y \rangle $ having the very same symmetry as $\varphi$. However, realistic microscopic interactions, e.g. ferromagnetic interorbital Hund's couplings, do not stabilize such a mean field order parameter $\eta$. In contrast, our vestigial $\varphi$ is a composite order parameter generated by a very different microscopic mechanism. In other words, fluctuations can stabilize unexpected AM phases at elevated temperatures. 

In this letter, we present a general phenomenological theory of `vestigial AM', which can be applied to the region of the phase diagram above the intersection of the AFM and OO critical lines where AFM and OO are absent, as shown in Fig.~\ref{fig:phasediagram_schematic}(a). There, due to the proximity to the critical points, large $M$ and $N$ fluctuations can produce a non-zero $\varphi$ with distinct experimental signatures.

\textit{Model.}
A system with the phase diagram in Fig.~\ref{fig:phasediagram_schematic}(a) can be described by the following Landau free energy:
\begin{equation}\label{eq:free-energy}
\begin{split}
    &F[M,N] =F_0+V, \\
    &F_0=\frac{1}{2}\sum_\mathbf{q}\left[\left(q^2+ r_{\text{AFM}}^{(0)}\right)|M_\mathbf{q}|^2 +\left(q^2+ r_{\text{OO}}^{(0)}\right)|N_\mathbf{q}|^2\right] \\ 
    &V =\int \left( \frac{u_1}{4} M^4 + \frac{u_2}{4}N^4 - \frac{v}{2} M^2N^2\right) \diff^2x.
\end{split}
\end{equation}
Here $\mathbf{q}$ is the small momentum deviation from the ordering wave vector $\mathbf{Q}$, and we have incorporated the effective masses into $M_\mathbf{q}$ and $N_\mathbf{q}$~\footnote{The effective masses are isotropic because $\mathbf{Q}$ hence the free energy is C$_4$-invariant and any C$_4$-invariant symmetric $2\times 2$ matrix is proportional to the identity matrix.}. The coupling constants in $V$ satisfy the conditions $u_1>0, u_1u_2 >v^2$ which ensure that the free energy is positive definite. Eq.~\eqref{eq:free-energy} can be derived from the underlying two- and three-orbital Hamiltonians, as is shown in the Supplemental Material (SM)~\cite{SM}. We note that our model reduces to that of Ref.~\cite{fernandes2012preemptive} in the limit of $u_1=u_2$ and $r_{\text{AFM}}^{(0)}=r_{\text{OO}}^{(0)}$, when $M, N$ become equivalent.

A second order mean-field (MF) transition to AFM or OO occurs when one of the susceptibilities $r_{\text{AFM}}^{(0)}$ or $r_{\text{OO}}^{(0)}$ becomes negative, and $\langle M\rangle$ or $\langle N \rangle$ becomes non-zero. Therefore, near the intersection point in Fig.~\ref{fig:phasediagram_schematic}(a), the MF phase diagram can be described using axes $r_{\text{AFM}}^{(0)}, r_{\text{OO}}^{(0)}$ which are small and are functions of $g, T$ as shown in Fig.~\ref{fig:phasediagram_schematic}(b). For example, at $r_{\text{AFM}}^{(0)},r_{\text{OO}}^{(0)}<0$ both AFM and OO exist and the system is in the conventional AM phase. We shall now show in the classical limit that for attractive interaction $v>0$, the composite order parameter $\varphi\propto \langle M N\rangle$ can become non-zero when $r_{\text{AFM}}^{(0)},r_{\text{OO}}^{(0)}>0$ due to fluctuations.

To take into account fluctuations of the order parameters, let us consider the partition function:
\begin{equation}
    Z = \int \Diff M \Diff N \exp(-F[M,N]/T),
\end{equation}
and perform the Hubbard-Stratonovich transformation to decouple the quartic interactions $V$ in the free energy~\eqref{eq:free-energy}:
\begin{equation}\label{eq:free-energy-effective}
\begin{split}
    F_{\text{eff}} =& F_0 + \int\diff^2x \bigg[-\frac{\psi_1^2}{4u_1}-\frac{\psi_2^2}{4u_2} +\frac{\varphi^2}{2v}\\
    &+ \frac{1}{2}\left( M^2\psi_1 + N^2\psi_2\right) - \varphi MN \bigg].
\end{split}
\end{equation}
The auxiliary boson fields $\psi_1,\psi_2,\varphi$ correspond to the following MF channels:
\begin{equation}\label{eq:MF-eq}
    \psi_1=u_1 \langle M^2 \rangle, \ \psi_2=u_2 \langle N^2 \rangle, \ \varphi = v\langle MN \rangle.
\end{equation}
This allows us to determine $\varphi$ and the vestigial AM phase in the saddle-point (or MF) approximation by solving Eq.~\eqref{eq:MF-eq}, i.e. integrating out the $M,N$ fields and minimizing the effective free energy.  We note that the $M^2N^2$ term can be also decoupled as $\psi_1\psi_2$; different choices of the decoupling schemes are studied in detail in Ref.~\cite{palle2026unbiased}, but here we decouple in $\varphi$ to obtain the vestigial phase.. Leaving the details of deriving and solving the resulting equations to the SM~\cite{SM}, in the main text we concentrate on the results. For this purpose, we define reduced quantities in units of the coupling constant $v$: the reduced coupling constants $\tilde{u}_i = u_i/v$, and similarly for a given quantity $f$ we define $f^* = 2f/\bar{v}$ where $\bar{v} = vT/2\pi$.

\begin{figure}
	\centering
	\includegraphics[width=\linewidth]{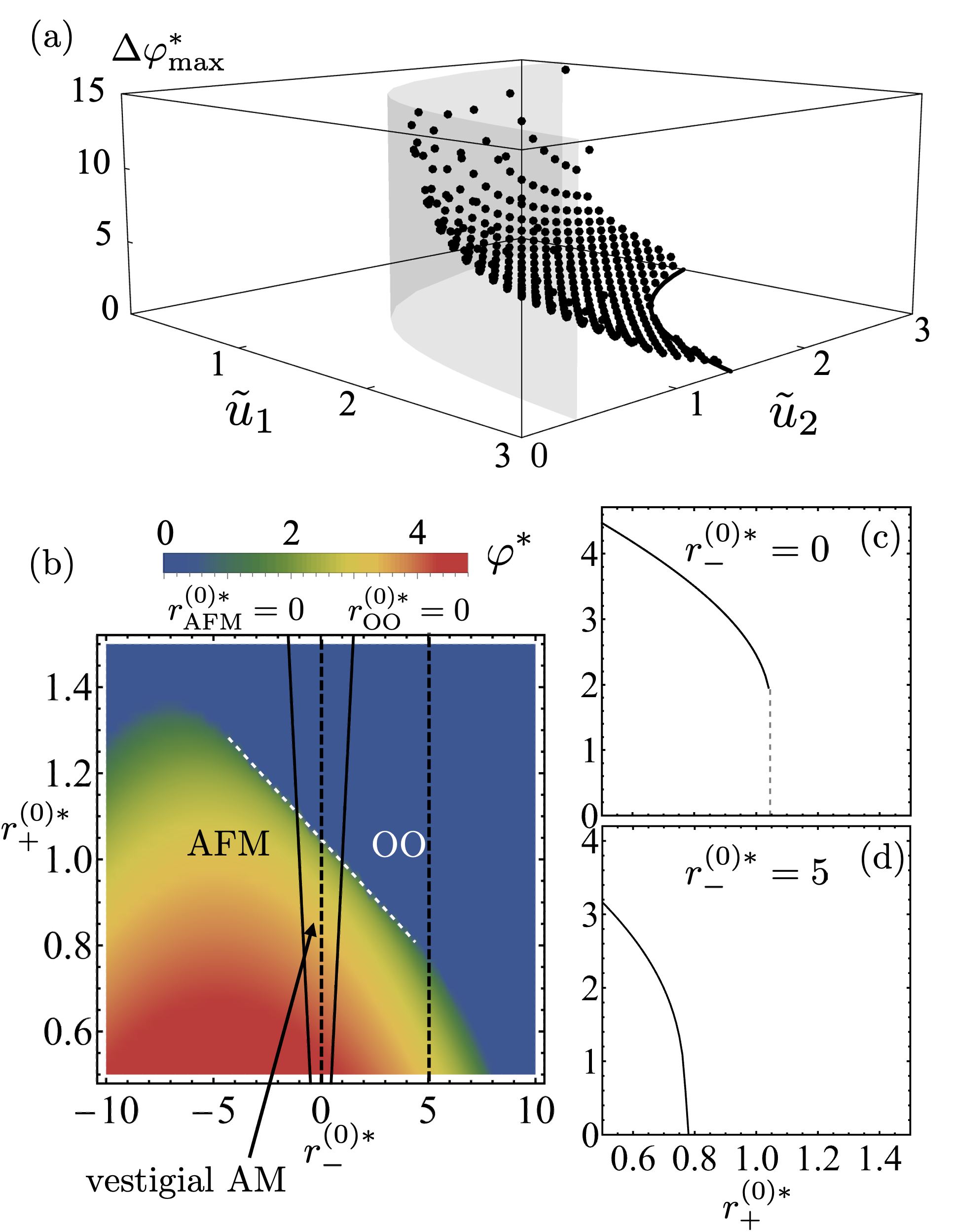}
	\caption{(a) 3D point plot of the largest possible first-order discontinuity $\Delta \varphi_{\text{max}}^*$ from the disordered phase into the vestigial AM phase, as a function of reduced coupling constants $\tilde{u}_i = u_i/v$. The gray vertical surface corresponds to the first of Eq.~\eqref{eq:criticallines} beyond which the system becomes unstable. The black solid line is given by the second of Eq.~\eqref{eq:criticallines} beyond which the transition is second order. (b) Phase diagram in $r_+^{(0)*},r_-^{(0)*}$ for $\tilde{u}_1= 1.6,\  \tilde{u}_2= 1.9$. $r_{\text{AFM}}^{(0)*}=0$ ($r_{\text{OO}}^{(0)*}=0$) is given by the left (right) black line; to its left (right) is the AFM (OO) phase. The region between the two black lines and the white dashed line is the vestigial AM phase. The transition is first order across the white dashed line which terminates at larger $|r_-^{(0)*}|$ into a second order transition; compare Fig.~\ref{fig:phasediagram_schematic}(b). Cuts at (c) $r_-^{(0)*}= 0$ and (d) $r_-^{(0)*}= 5$ respectively. In (c) the first order transition is marked by a dashed gray vertical line.}
	\label{fig:phasediagram}
\end{figure}

\textit{Phase diagram.}
For given coupling constants $\tilde{u}_1,\tilde{u}_2$, the phase diagram is determined by the susceptibilities which are governed by experimental conditions such as temperature. Therefore let us first consider the parameter space of $\tilde{u}_1$ and $\tilde{u}_2$ before discussing the phase diagram for specific $\tilde{u}_1,\tilde{u}_2$. We find three regions separated by the following two curves:
\begin{equation}\label{eq:criticallines}
  \tilde{u}_1\tilde{u}_2=1, \ 2\tilde{u}_1\tilde{u}_2- \tilde{u}_1-\tilde{u}_2 = 4.  
\end{equation}
The inequality $\tilde{u}_1\tilde{u}_2<1$ corresponds to $u_1u_2<v^2$ and the free energy is not bounded from below: the system is unstable. Between the regions given by the two curves in Eq.~\eqref{eq:criticallines}, the phase boundary separating the vestigial AM and the high-temperature disordered phases is generally second order. However, there can be critical lines of first order transitions in the phase diagram across which $\varphi$ acquires a discontinuous jump as the temperature decreases; this is shown schematically as black dashed curves in Fig.~\ref{fig:phasediagram_schematic}. The MF equations set an upper limit on the first order discontinuity $\Delta \varphi_{\text{max}}^*$ as a function of $\tilde{u}_1,\tilde{u}_2$~\cite{SM}, which we plot in Fig.~\ref{fig:phasediagram}(a). Towards the curve $\tilde{u}_1\tilde{u}_2= 1$, $\Delta \varphi_{\text{max}}^*$ tends to infinity. At $2\tilde{u}_1\tilde{u}_2- \tilde{u}_1-\tilde{u}_2 >4$ the vestigial AM transition is always second order in the phase diagram and $\Delta \varphi_{\text{max}}^*=0$ as a result.

We now study the system for specific values of $\tilde{u}_1,\tilde{u}_2$ and present the phase diagrams in the bare variables $r_\pm^{(0)*}= (r_{\text{AFM}}^{(0)*}\pm r_{\text{OO}}^{(0)*})/2$. However, fluctuations introduce a logarithmic correction $\log(2\Lambda^2/\bar{v})$ into the MF equations, where $\Lambda$ is the momentum cut-off. This term cannot be incorporated into one of the bare parameters but we shall take it to be zero: as is discussed in the SM~\cite{SM}, a non-zero correction does not change qualitatively our conclusions. As an example, let us take $\tilde{u}_1= 1.6,\  \tilde{u}_2= 1.9$. The phase diagram is given in Fig.~\ref{fig:phasediagram}(b) where we also show the lines $r_{\text{AFM}}^{(0)*}=0$ and $r_{\text{OO}}^{(0)*}=0$ explicitly as solid black lines. Across the white dashed line the transition is first order. The line terminates at sufficiently large $|r_-^{(0)*}|$ and the the transition becomes second order. In our case, the transition is first order throughout the region $r_{\text{AFM}}^{(0)*}, r_{\text{OO}}^{(0)*}>0$ at which the vestigial AM phase is located. We also display two cuts at $r_-^{(0)*}= 0$ and $r_-^{(0)*}= 5$ in Fig.~\ref{fig:phasediagram}(c)-(d) to demonstrate the nature of the transition. Finally, in the SM we also verify that for the second order transition, the critical exponent in $\varphi^*\propto (T_c-T)^{\alpha}$ has the MF value $\alpha = 1/2$~\cite{SM}.

\begin{figure}
    \centering
    \includegraphics[width=\linewidth]{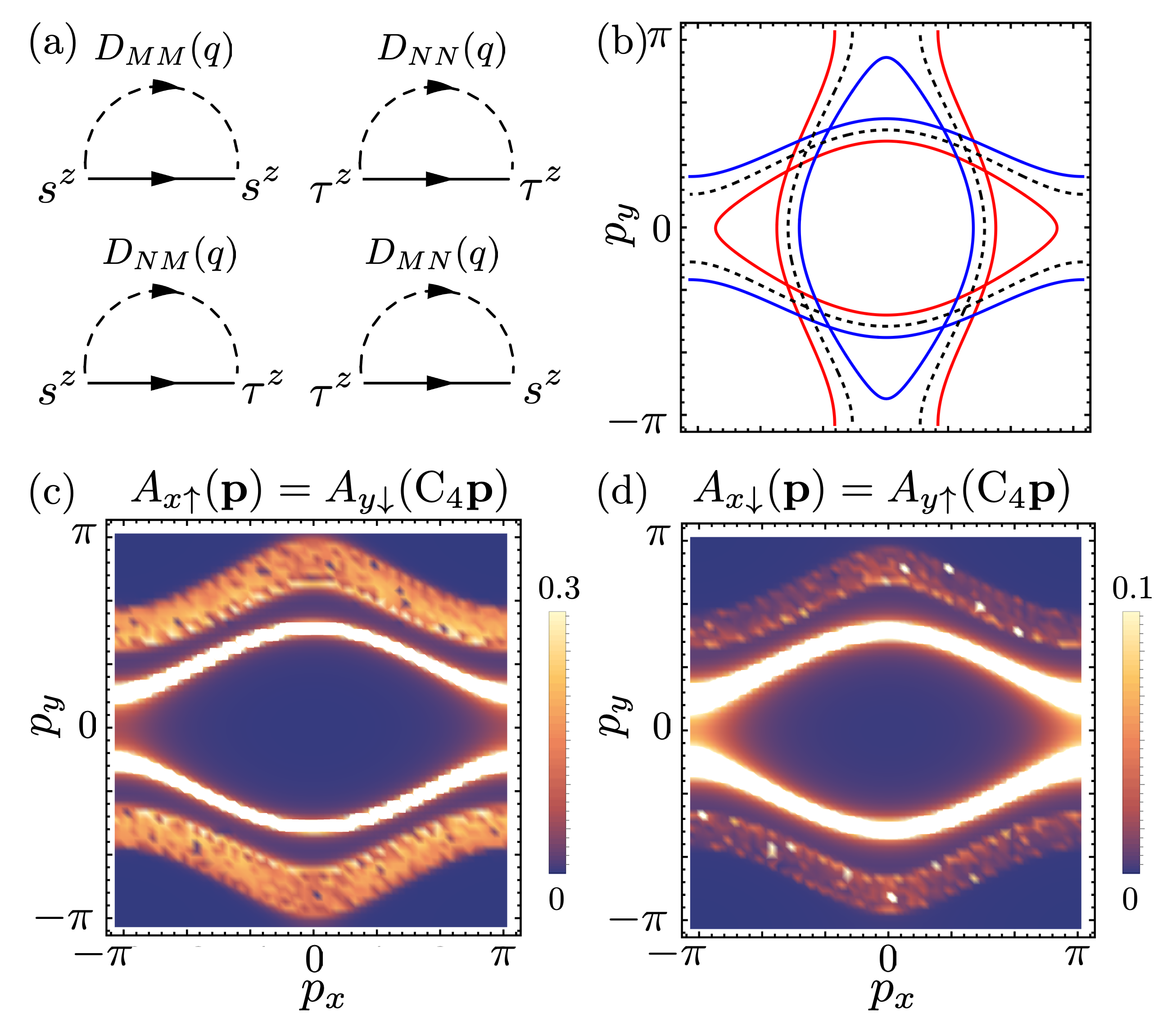}
    \caption{(a) One-loop self-energy diagrams where the $s^z$ and $\tau^z$ vertices in Eq.~\eqref{eq:el-boson-vertex} are shown explicitly.  (b) The spin split Fermi surface given by $H_0$ in Eq.~\eqref{eq:model-electron} and $\Re \Sigma_{\text{AM}} = 0.5 t s^z\tau^z$ where $t$ is the nearest-neighbour hopping amplitude. Red is spin up and blue spin down. The black dashed lines correspond to no AM splittings. Spectral function for (c) spin-up and (d) spin-down electrons in orbital $x$ with $\Re \Sigma_{\text{AM}} = 0$ at $T=0.05t$. The spectral functions for $y$-orbitals are related by a C$_4$ rotation and opposite spins. The inner contour corresponds to the $x$-orbital Fermi surface shown in black dashed line in (b). The outer ring is given by the inner ring shifted by $\mathbf{Q}=(\pi,\pi)$. The isolated bright spots on the outer ring are numerical defects from evaluating the self-energy.}
    \label{fig:spectralweight}
\end{figure}

\textit{Spectral functions.}
Next, we discuss the unique properties and experimental signatures of the vestigial AM. 
As a consequence of the combined rotation and time-reversal symmetry of the AM, the Fermi surface is expected to be spin-split and the splitting can become comparable to the Fermi energy~\cite{hayami2019momentum,yuan2020giant,ahn2019afm,smejkal2020crystal,smejkal2022am1}. However, the vestigial AM order parameter $\varphi$ arises from interactions between $M$ and $N$ and does not couple to the electron densities directly. It might then seem that the on-set of vestigial AM will have no effect on the electron spectra. But as seen from Eq.~\eqref{eq:free-energy-effective}, $\varphi$ couples to the product of $M$ and $N$. Therefore, the vestigal AM appears in the self-energy corrections via the electron couplings
\begin{equation}\label{eq:el-boson-vertex}
    V_{\text{el-b}} = \sum_{\mathbf{p},\mathbf{q}}\left(M_\mathbf{q} \Psi_{\mathbf{p}+\mathbf{Q}+\mathbf{q}}^\dagger s^z \Psi_\mathbf{p} +  N_\mathbf{q}\Psi_{\mathbf{p}+\mathbf{Q}+\mathbf{q}}^\dagger \tau^z \Psi_\mathbf{p}\right).
\end{equation}
Note the additional momentum $\mathbf{Q}$ carried by the bosonic fluctuations $M_\mathbf{q}$ and $N_\mathbf{q}$. The corresponding one-loop diagrams are shown in Fig.~\ref{fig:spectralweight}(a). In the absence of vestigial AM, $M$ and $N$ fields are decoupled on the quadratic level in Eq.~\eqref{eq:free-energy-effective}, and only the upper two diagrams contribute with the bosonic propagators $D_{MM}(x)= \langle \text{T}_\tau \{ M(x)M(0) \} \rangle $ and $D_{NN}= \langle \text{T}_\tau \{ N(x)N(0) \} \rangle $; here $\tau$ is the imaginary time and $x=(\tau,\mathbf{r})$. A finite $\varphi$ couples $M$ and $N$ on the quadratic level, which results in a non-zero off-diagonal $D_{MN}(x) = \langle \text{T}_\tau \{ M(x)N(0) \} \rangle$. Consequently, the vestigial AM order gives a self-energy contribution $\tau^z  s^z \Sigma_{\text{AM}}(\varepsilon,\mathbf{p})$ corresponding to the lower two diagrams in Fig.~\ref{fig:spectralweight}(a). This term is C$_4$T-symmetric because C$_4$ exchanges orbitals and $\tau^z\rightarrow -\tau^z$, and under time-reversal $s^z\rightarrow -s^z$. 

Via the self energy the vestigial AM manifests in two distinct ways. First, at small $\omega$ and near the Fermi momentum $\mathbf{p}_F$, $\Re \Sigma_{\text{AM}}$ gives a spin- and orbital-dependent constant energy correction, which is responsible for the spin-split AM Fermi surface akin to the MF $\eta$ of Ref.\cite{giuli2025altermagnetism}. Second, and more interestingly, $\Im \Sigma_{\text{AM}}$ gives a spin- and orbital-dependent contribution to the {\it quasiparticle decay rate} $\gamma(\omega,\mathbf{p})$! In particular, this means that the Fermi surfaces will have unique spin dependent broadening with AM symmetires~\footnote{It might appear that $\gamma(\omega,\mathbf{p})$ can be negative due to the $s^z\tau^z$ vertex in the lower two diagrams in Fig.~\ref{fig:spectralweight}(a). However, after summing over all diagrams in Fig.~\eqref{fig:spectralweight}(a) $\gamma$ should always be positive because it can be equivalently given by Fermi's golden rule $\gamma \propto \sum_f |(V_{\text{el-b}})_{fi}|^2\geq 0$. }. 

To illustrate these points in a concrete system, let us consider the minimal two-orbital electron model of Ref.~\cite{leeb2024spontaneous} with the Hamiltonian $H = H_0 +V_{\text{el-b}}$ where:
\begin{equation}\label{eq:model-electron}
 H_0 = -2t (\cos p_x +\cos p_y) +2 \delta t (\cos p_x- \cos p_y) \tau^z.
\end{equation}
For simplicity only the nearest-neighbour intra-orbital transitions are included and $\delta t$ describes the anisotropic overlap of $x,y$ orbitals along the $x$- and $y$-axes as shown in Fig.~\ref{fig:phasediagram_schematic}(a). We demonstrate the effect of a finite $\varphi$ on the electron spectra and fix $ \Re \Sigma_{\text{AM}} = 0.5 t$. The spin-split C$_4$T-symmetric Fermi surfaces at $\delta t =0.4 t$ and $\mu= -1.2 t$ are shown in Fig.~\eqref{fig:spectralweight}(b). Note that since the vestigial AM has zero ordering wave vector, no Brillouin zone (BZ) backfolding is needed, unlike the conventional AM at lower temperatures. 

Next, we concentrate on the quasiparticle decay effect by computing the zero frequency spectral function as a function of momenta at $T=0.05t$ for electrons with orbital $\lambda = x,y$ and spin $s=\uparrow, \downarrow$:
\begin{equation}\label{eq:spectral-function}
    A_{\lambda s}(\mathbf{p}) = -\frac{\Im \Sigma^R_{\lambda s}(0,\mathbf{p})}{[\varepsilon_{\lambda s}(\mathbf{p})-\mu+ \Re \Sigma_{\lambda s}^R(0,\mathbf{p})]^2 + [\Im \Sigma_{\lambda s}^R(0,\mathbf{p})]^2}.
\end{equation}
$\varepsilon_{\lambda s}(\mathbf{p})$ is the electron dispersion for spin $s$ and orbital $\lambda$ and $\Sigma_{\lambda s}^R(\omega,\mathbf{p})$ is the retarded self-energy for the electron band with given $\lambda$ and $s$. For numerical simulations of $\Im \Sigma_{\lambda s}^R(\omega,\mathbf{p})$, the boson propagators are given by the following Lagrangian:
\begin{equation}\label{eq:boson-Lagrangian}
\begin{split}
    \mathcal{L} =\int \diff \tau  \bigg\{ \frac{1}{2}\left[(\p_\tau M)^2+ (\p_\tau N)^2\right] -\varphi_0 M N+F_0\bigg\},
\end{split}
\end{equation}
where we take $r_{\text{AFM}}^{(0)} = 0.1 t, \ r_{\text{OO}}^{(0)} = 0.2 t$. The vestigial AM phase is characterized by $\varphi_0 = \langle \varphi \rangle = 0.2 t$. The details of the analytical and numerical evaluations are given in the SM~\cite{SM}, and the results for $\lambda= x$ and $s=\uparrow,\downarrow$ are presented in Fig.~\ref{fig:spectralweight}(c)-(d). In the numerical simulation, we omit AM splitting by setting $\Re \Sigma_{\lambda s} = 0$ in Eq.\eqref{eq:spectral-function} to focus on features due to the quasiparticle lifetime alone. To take into account broadening due to phonons and disorder we also fix $\Im \Sigma\rightarrow \Im \Sigma+ 0.01 t$. Generally, $A_{\lambda s}$ is concentrated near the Fermi surface, and a broader region appears which is  shifted from the Fermi surface by $\mathbf{Q}$ due to the large momentum transfer $\mathbf{Q}$ from scattering by the bosons. Moreover, $A_{x\uparrow}$ and $A_{x\downarrow}$ have different intensities despite the spin-degenerate Fermi surface. The results for the $y$-orbitals are related by a C$_4$ rotation and changing the spin direction, i.e. time-reversal. Therefore, we have shown that vestigial AM can manifest as a pure lifetime effect if the fluctuations dominate the imaginary part of the self-energy. We note that it can be shown that the dominant scattering region is given by thermal broadening $|\varepsilon_\lambda (\mathbf{p}+\mathbf{Q})-\mu |\sim T$ ~\cite{SM}. The shifted `shadow Fermi surface' in the spectral function is known for electronic models of quantum criticality adjacent to AFM order; see Ref.~\cite{abanov2003quantum} and references therein.

\textit{Discussions and Outlook.}
In this letter, we showed that the thermal melting of AM phases which arise from the coexistence of distinct order parameters, e.g. AFM with OO, can give rise to new composite phases. The resulting vestigial AM order appears at elevated temperatures above the low temperature ordered phases. Depending on the interaction parameters, the transition from the high-temperature disordered phase into the vestigial phase can be of first or second order. For parameters lying between the curves of Eq.~\eqref{eq:criticallines}, there is a first order critical line which terminates into a second order transition, although AFM and OO may set in before the termination. We further studied the effect of vestigial AM on the electron spectra using a simplified two-orbital model, and showed that the zero frequency spectral functions are spin- and orbital-dependent and C$_4$T-symmetric. A unique signature would be a anisotropic quasiparticle damping rate with AM symmetries. 

There are other unconventional magnetic systems beyond AMs with multiple phases, e.g. odd wave $p$-wave magnets can emerge from the coeistence of AFM and loop current orders~\cite{leeb2026collinear}. Determining the nature and symmetry of the possible vestigial phases in these systems remains an interesting open question. 

\vspace{1cm}
Note added: We recently learned of a related vestigial AM study for the spin- and orbital-degenerate Kugel-Khomskii model~\cite{valiulin2026altermagnetism}.

\vspace{1cm}
\begin{acknowledgments}

P.R. thanks J. Habel and M. Ciarchi for interesting discussions and helpful comments on the manuscript. J.K. thanks V. Leeb for helpful discussions and related collaborations. J.K. acknowledges support from the Deutsche Forschungsgemeinschaft (DFG, German Research Foundation) under grants TRR 360 - 492547816, KN1254/1-2, KN1254/2-1 and under Germany’s Excellence Strategy EXC-2111-390814868. P.R. and J.K. acknowledge support from the Munich Quantum Valley, which is supported by the Bavarian state government with funds from the High-tech Agenda Bayern Plus. J.K. further acknowledges support from the Imperial-TUM flagship partnership and the Keck foundation.

\end{acknowledgments}

\vspace{1cm}
The code that numerically solves the MF equations and produces the results in this letter is available at \cite{code} upon reasonable request.

\nobalance 


%

\end{document}


\title{Supplemental Material for `Vestigial altermagnetism'}

\author{Peng Rao}
\affiliation{Technical University of Munich, TUM School of Natural Sciences, Physics Department, 85748 Garching, Germany}
\author{Johannes Knolle}
\affiliation{Technical University of Munich, TUM School of Natural Sciences, Physics Department, 85748 Garching, Germany}
\affiliation{Munich Center for Quantum Science and Technology (MCQST), Schellingstr. 4, 80799 München, Germany}
\date{\today}

\date{\today}
\begin{abstract}
\tableofcontents
\end{abstract}

\maketitle




\section{The effective free energy}

Let us start from the partition function as a functional integral in imaginary time $\tau$:
\begin{equation}
    Z = \int \Diff M \Diff N \exp (-S), \ S = \int_0^{1/T} \mathcal{L}\diff \tau \diff^2x,
\end{equation}
with the Lagrangian density:
\begin{equation}\label{eq:SM:Lagrangian}
    \mathcal{L} =\mathcal{K}+ \frac{1}{2}\sum_i\left[\frac{1}{2m_1}(\nabla M)^2+  r_{\text{AFM}}^{(0)} M^2+\frac{1}{2m_2}(\nabla N)^2+ r_{\text{OO}}^{(0)} N^2\right] + \frac{u_1}{4} M^4 + \frac{u_2}{4}N^4 - \frac{v}{2} M^2N^2.
\end{equation}
For completeness, we include the effective masses $m_1,m_2$ for the $M,N$ fields for now. $\mathcal{K}$ is the kinetic term for $M$ and $N$, which we shall neglect by taking the classical limit. Then the integration over $\tau$ can be performed directly and we get Eq.~(\ref*{M:eq:free-energy}) in the main text. 

For notation convenience, the $M$ and $N$ fields are written as components of a single boson doublet $\phi = (M,N)$ with $r_{10}= r_{\text{AFM}}^{(0)}, \ r_{20} = r_{\text{OO}}^{(0)}$. The standard Hubbard-Stratonovich decoupling of the quartic terms in Eq.~\eqref{eq:SM:Lagrangian} gives:
\begin{equation}\label{eq:SM:Lagrangian-1}
    \mathcal{L} = -\frac{\psi_1^2}{4u_1}-\frac{\psi_2^2}{4u_2} +\frac{\varphi^2 }{2v} +\frac{1}{2}\sum_i \left(r_{i0} +\psi_i+ \frac{q^2}{2m_i}\right) \phi_i^2-\varphi \phi_1\phi_2,
\end{equation}
which in the saddle-point approximation (MF) with resepct to $\psi_i,\varphi$ gives:
\begin{equation}
    \psi_1 =u_1\langle  M^2\rangle, \ \psi_2 = u_2 \langle  N^2\rangle, \ \varphi = v \langle M N \rangle.
\end{equation}

Finally, we comment on some subtleties of decoupling the repulsive interactions given by $u_1,u_2$, as the $-\psi_i^2/4u_i$ term in the Lagrangian is not positive definite. Let us consider the $u_1$ term as an example. Following standard procedure, we introduce the real auxiliary boson field $\psi_1$ into the partition function and perform a change of variables $\psi_1 \rightarrow \psi_1 + i u_1M^2$:
\begin{equation}
    \int \Diff \psi_1 \exp\bigg(- \int \frac{\psi_1^2 }{4u_1}\diff^2x\diff \tau\bigg) e^{-(u_1/4) \int M^4\diff^2 x\diff \tau} \rightarrow \int \Diff \psi_1 \exp\bigg[-\int\bigg(\frac{\psi_1^2}{4u_1} +\frac{i}{2}\psi_1 M^2\bigg)\diff^2x\diff \tau\bigg].
\end{equation}
Contrary to the case of attractive interactions, the action is now complex. We then formally make the action real by $\psi_1 \rightarrow- i\psi_1$:
\begin{equation}
    \int_{-i\infty -u_1M^2}^{i\infty-u_1M^2} \Diff \psi_1 \exp\bigg[\int\bigg(\frac{\psi_1^2}{4u_1} -\frac{1}{2}\psi_1 M^2\bigg)\diff^2 x\diff \tau\bigg].
\end{equation}
This gives the $\psi_1$ terms in Eq.~\eqref{eq:SM:Lagrangian-1} although the integration over $\psi_1$ is now taken over a vertical line in the complex plane. In taking the saddle-point approximation to obtain a real MF solution, the integration contour is displaced onto the real axis at which the saddle-point is located. The result is real, as should the partition function.

\section{MF equations}\label{sec:SM:MF}

We now integrate over $\phi$ using the Lagrangian~\eqref{eq:SM:Lagrangian-1} to obtain the effective action:
\begin{equation}
    \mathcal{L}_{\text{eff}} =\int \bigg(-\sum_{i=1}^2\frac{\psi_i^2}{4u_i } +\frac{\varphi^2}{2v}\bigg)\ \diff^2x +\frac{T}{2}\sum_\mathbf{q}\tr \log \begin{pmatrix}
      \frac{q^2}{2m_1} +r_1  & -\varphi\\ 
      -\varphi & \frac{q^2}{2m_2} +r_2
    \end{pmatrix},
\end{equation}
where $r_i = r_{i0}+\psi_i$ are the renormalized susceptibilities. In the last term, the summation over Matsubara frequency has been replaced by multiplication by $T$ in the classical limit. Using the identity $\tr \log A = \log \det A$ and varying $S_{\text{eff}}$ with respect to $\psi_1,\psi_2,\varphi$ give the MF equations:
\begin{align}\label{eq:SM:MF-1}
    r_1 &=r_{10}+ u_1T\int \frac{\diff^2q}{(2\pi)^2} \frac{r_2 + \frac{q^2}{2m_2}}{\left(r_1+ \frac{q^2}{2m_1}\right)\left(r_2 + \frac{q^2}{2m_2}\right)-\varphi^2}, \\ 
   r_2 &=r_{20}+ u_2T\int \frac{\diff^2q}{(2\pi)^2} \frac{r_1 + \frac{q^2}{2m_1}}{\left(r_1+ \frac{q^2}{2m_1}\right)\left(r_2 + \frac{q^2}{2m_2}\right)-\varphi^2}, \\
   \varphi& =vT\int \frac{\diff^2q}{(2\pi)^2} \frac{\varphi}{\left(r_1+ \frac{q^2}{2m_1}\right)\left(r_2 + \frac{q^2}{2m_2}\right)-\varphi^2}.
\end{align}
The MF equations~\eqref{eq:SM:MF-1} reduce to those in Ref.~\cite{fernandes2012preemptive} in the limit of $u_1=u_2,r_{10}=r_{20},  \psi_1=\psi_2$. In fact in this limit, Eq.~\eqref{eq:SM:Lagrangian-1} becomes the Lagrangian in Ref.~\cite{fernandes2012preemptive} after the substitution $M = (\Delta_X +\Delta_N)/\sqrt{2}$ and $N = (\Delta_X -\Delta_N)/\sqrt{2}$.

The integrations in Eq.~\eqref{eq:SM:MF-1} can be performed directly. In terms of quantities:
\begin{equation}
    r_+ = m_1r_1 +m_2r_2, \ r_- = m_1r_1 -m_2r_2, \ c = \sqrt{r_-^2+4m_1m_2\varphi^2},
\end{equation}
and similarly for $r^{(0)}_\pm$ in terms of $r_{i0}$, the results for Eq.~\eqref{eq:SM:MF-1} can be represented as:
\begin{equation}
\begin{split}
r_1 &= r_{10} - \frac{\bar{u}_1m_1 r_-}{2c}\log \frac{r_++c}{r_--c} +\frac{\bar{u}_1m_1}{2} \log \frac{\Lambda^4}{r_+^2-c^2},\\ 
r_{2} &= r_{20} + \frac{\bar{u}_2m_2 r_-}{2c}\log \frac{r_++c}{r_--c} +\frac{\bar{u}_2m_2}{2} \log \frac{\Lambda^4}{r_+^2-c^2}, \\ 
r_+ &= c\coth (2c/\bar{v}).
\end{split}
\end{equation}
Here $ \bar{u}_i = uT/2\pi$ and $\Lambda$ is the momentum cut-off which keeps the logarithm dimensionless. In what follows it is more convenient to consider the MF equations for $r_\pm$ 
\begin{align}\label{eq:SM:MF-2}
    r_+ =& r_{+}^{(0)} -\frac{r_-}{2c} (m_1^2\bar{u}_1 - m_2^2\bar{u}_2) \log\frac{r_++c}{r_+-c} +\frac{\bar{u}_1m_1^2 +\bar{u}_2m_2^2  }{2}\log \frac{\Lambda^4}{r_+^2-c^2}. \\
    r_- =&r_{-}^{(0)} -\frac{r_-}{2c} (m_1^2\bar{u}_1 + m_2^2\bar{u}_2) \log\frac{r_++c}{r_+-c} +\frac{\bar{u}_1m_1^2 -\bar{u}_2m_2^2  }{2}\log \frac{\Lambda^4}{r_+^2-c^2}.
\end{align}
Here we note that the coupling constants $\bar{u}_i,\bar{v}$ only appear in the combination $m_i^2\bar{u}_i, \ m_1m_2\bar{v}$. Thus we shall make the following substitution in Eq.~\eqref{eq:SM:MF-2}:
\begin{equation}
    m_i^2\bar{u}_i  \rightarrow\frac{\bar{u}_i}{4}, \ m_1m_2\bar{v} \rightarrow \frac{\bar{v}}{4}.
\end{equation}
This is equivalent to making the substitution $M\rightarrow \sqrt{2m_1}M,  N \rightarrow \sqrt{2m_2}N$ in Eq.~\eqref{eq:SM:Lagrangian} which removes the effective mass as we have done in Eq.~(\ref*{M:eq:free-energy}). Equivalently, we can neglect the effective mass by taking $m_i=1/2$ which we shall do from this point on.

Returning to the MF equations, let us define reduced quantities in relation to $\bar{v}$:
\begin{equation}
    \tilde{u}_i = \frac{\bar{u}_i}{\bar{v}}, \ r_+^{*} = \frac{2r_+}{\bar{v}}, \ r_-^{*}=\frac{2r}{\bar{v}}, \ c^* =\frac{2c}{\bar{v}}.
\end{equation}
and Eq.~\eqref{eq:SM:MF-2} can be written as:
\begin{align}\label{eq:SM:MF-3}
    r_+^{*} &= r_+^{(0)*} -\frac{r_-^{*}}{4c^*} (\tilde{u}_1-\tilde{u}_2) \log\frac{r_+^{*}+c^*}{r_+^{*}-c^*} +\frac{\tilde{u}_1+\tilde{u}_2  }{2}\log \frac{2\Lambda^2/\bar{v}}{\sqrt{r_+^{*2}-c^{*2}}}, \\
    r_-^{*} &=r_-^{(0)*} -\frac{r_-^{*}}{4c^*} (\tilde{u}_1+\tilde{u}_2) \log\frac{r_+^{*}+c^*}{r_+^{*}-c^*} +\frac{\tilde{u}_1-\tilde{u}_2  }{2}\log \frac{2\Lambda^2/\bar{v}}{\sqrt{r_+^{*2}-c^{*2}}}, \\ 
    r_+^{*} &= c^*\coth c^*.  
\end{align}
Using the following identities due to the third equation:
\begin{equation}
    \log\frac{r_+^{*}+c^*}{r_+^{*}-c^*}  = 2c^*, \ r_+^{*2} - c^{*2} = \frac{c^{*2}}{\sinh^2c^*},
\end{equation}
we first solve for $r_-^{*}$ in the second equation:
\begin{equation}\label{eq:SM:MF-reduced-1}
    r_-^{*} = \frac{2r_-^{(0)*}}{2+\tilde{u}_1+\tilde{u}_2} + \frac{\tilde{u}_1-\tilde{u}_2}{2+\tilde{u}_1+\tilde{u}_2}\left(\log \frac{2\Lambda^2}{\bar{v}} - \log \frac{c^*}{ \sinh c^* }\right),
\end{equation}
The result is then substituted into the first equation to obtain:
\begin{equation}\label{eq:SM:MF-reduced}
\begin{split}
    c^*\coth c^*+\frac{1}{2}\left[\tilde{u}_1+\tilde{u}_2 -\frac{(\tilde{u}_1-\tilde{u}_2)^2}{2+\tilde{u}_1+\tilde{u}_2}  \right]\log \frac{c^*}{\sinh c^*} =r_+^{(0)*} -\frac{r_-^{(0)*}(\tilde{u}_1-\tilde{u}_2)}{2+\tilde{u}_1+\tilde{u}_2}+\frac{1}{2}\left[\tilde{u}_1+\tilde{u}_2 -\frac{(\tilde{u}_1-\tilde{u}_2)^2}{2+\tilde{u}_1+\tilde{u}_2}  \right]\log \frac{2\Lambda^2 }{\bar{v}}.
\end{split}
\end{equation}
We eliminate the $\log(2\Lambda^2/\bar{v})$ term in Eq.~\eqref{eq:SM:MF-reduced-1} using Eq.~\eqref{eq:SM:MF-reduced}:
\begin{equation}\label{eq:SM:MF-reduced-2}
    r_-^{*} =\frac{(\tilde{u}_1+\tilde{u}_2)r_-^{(0)*}+(\tilde{u}_1-\tilde{u}_2)\left(c^*\coth c^*-r_+^{(0)*} \right)}{\tilde{u}_1+\tilde{u}_2+2\tilde{u}_1\tilde{u}_2}.
\end{equation}
Finally the reduced order paramter $\varphi^*$ is related to $c^*$ by:
\begin{equation}\label{eq:SM:order-parameter}
    \varphi^* =\sqrt{c^{*2} - r_-^{*2}}\leq c^*.
\end{equation}
Eqs.~\eqref{eq:SM:MF-reduced}, \eqref{eq:SM:MF-reduced-2} and \eqref{eq:SM:order-parameter} determine the MF behavior of the vestigial AM order.

\begin{figure}
    \centering
    \includegraphics[width=0.8\linewidth]{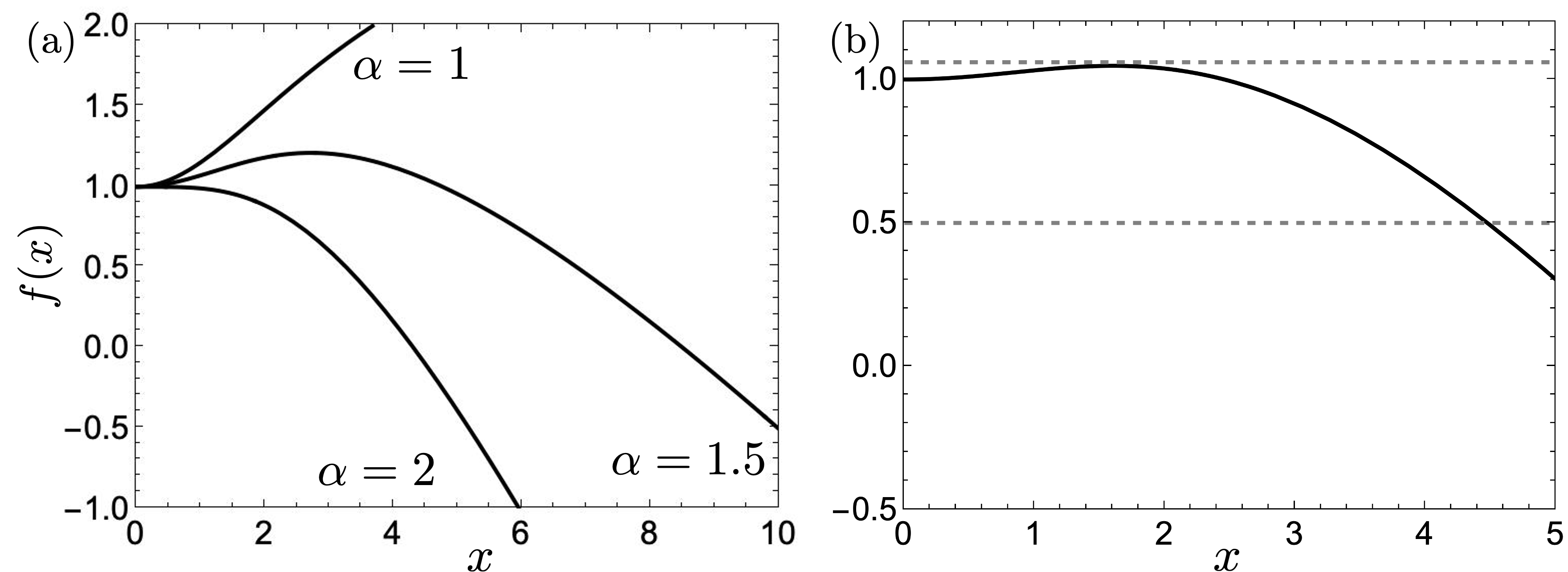}
    \caption{(a) The function $f(x)$ in Eq.~\eqref{eq:SM:MF-reduced-3}  plotted for $\alpha = 1, 1.5, 2$. For $\alpha \leq 1$, $f(x)$ increases monotonously. For $\alpha>1$, the local maxima decreases with increasing $\alpha$ and at $\alpha>2$ it is located at the origin only. (b) The function $f(x)$ for $\tilde{u}_1=1.6,\tilde{u}_2=1.9$. The vertical lines correspond to $f(x)= 0.5, 1.06$. At $f(x) \approx 1.06$ a solution $x$ for Eq.~\eqref{eq:SM:MF-reduced-3} first emerges.  }
    \label{fig:MF}
\end{figure}

\section{Phase diagram}
Let us now examine Eq.~\eqref{eq:SM:MF-reduced} more closely. The equation can be written in the form:
\begin{equation}\label{eq:SM:MF-reduced-3}
   f(c^*) = \bar{r}_0,  \  f(x) = x\coth x + \alpha \log \frac{x}{\sinh x} 
\end{equation}
with the constants:
\begin{align}
    \alpha = \frac{1}{2}\left[\tilde{u}_1+\tilde{u}_2 -\frac{(\tilde{u}_1-\tilde{u}_2)^2}{2+\tilde{u}_1+\tilde{u}_2}  \right], \
    \bar{r}_0 = r_+^{(0)*} -\frac{r_-^{(0)*}(\tilde{u}_1-\tilde{u}_2)}{2+\tilde{u}_1+\tilde{u}_2}+\alpha\log \frac{2\Lambda^2 }{\bar{v}}.
\end{align}
Thus the solution is determined by the intersection of the function $f(x)$ with the horizontal line $\bar{r}_0$. Asymptotically, $f(x) \sim(1-\alpha) x$ at large positive $x$ and at small $x$, $f(x)\sim  1-(\alpha-2)x^2/6$. It can be shown that for $\alpha<1$, $f(x)$ increases monotonously with positive $x$. For $1<\alpha<2$, $f(x)$ increases at small positive $x$ and reaches a maximum before decreasing linearly at large $x$. For $\alpha>2$, $f(x)$ decreases monotonously with $x$. The three behaviors of $f(x)$ are shown in Fig.~\ref{fig:MF}(a).

If $\alpha<1$, there is always a solution corresponding to arbitrarily large $\bar{r}_0$ and therefore $r_+^{(0)*}$, which is clearly unphysical. In fact, the condition $\alpha<1$ corresponds to $\tilde{u}_1\tilde{u}_2<1$ which is the same as the positive definiteness condition $u_1u_2>v^2$ for the free energy. When $1<\alpha<2$, upon decreasing $\bar{r}_0$ a MF solution emerges at non-zero $c^*_{\text{cr}}$ when $\bar{r}_0$ first reaches the local maximum. This corresponds to the appearance of an inflexion point in the free energy. Upon further decreasing $\bar{r}_0$, there are two solutions $c^*_1<c^*_{\text{cr}}< c^*_2$ where clearly $c^*_1$ corresponds to a local maxima of the free energy and $c^*_2$ the minima. As $\alpha\rightarrow 1$, $c^*_{\text{cr}}$ tends to infinity. This is demonstrated in Fig.~\ref{fig:MF}(b). Lastly at $\alpha>2$, a MF solution for $c^*$ first emerges at $c^*_{\text{cr}}=0$ which then increases with decreasing $\bar{r}_0$. In summary, there are two critical lines $\alpha=1, \alpha=2$ between which the transition for $c^*$ (not $\varphi^*$!) is first order and the transition is second order at $\alpha>2$. This gives:
\begin{equation}\label{eq:SM:criticallines}
   \alpha=1 \rightarrow\tilde{u}_1\tilde{u}_2=1, \ \alpha=2\rightarrow 2\tilde{u}_1\tilde{u}_2- \tilde{u}_1-\tilde{u}_2 = 4,
\end{equation}
which are precisely the critical lines in the main text. 

However, as already noticed the vestigial order parameter is $\varphi^*$ given by Eq.~\eqref{eq:SM:order-parameter}. Therefore, in Fig.~\ref*{M:fig:phasediagram} we also plot $c^*_{\text{cr}}$ as the largest possible discontinuity across the transition point $\Delta \varphi^*_{\text{max}}$, as a function of $\tilde{u}_1,\tilde{u}_2$. This is accomplished by numerically finding the value $c^*_{\text{cr}}$ at the local maxima of the left hand side of Eq.~\eqref{eq:SM:MF-reduced-3}, which only depends on $\tilde{u}_1,\tilde{u}_2$.

Regarding then the transition to the vestigial AM order, there are two possibilities. First, $c^*_{\text{cr}}>r_-^{*}$ which corresponds to a non-zero $\varphi^*_{\text{cr}}$. In this case, there is a first order transition to the vestigial AM. On the other hand, if $c^*_{\text{cr}}<r_-^{*}$ then $\varphi^*$ must be taken zero. In this case, a physical solution for $\varphi^*$ only emerges when $c^*_2 \geq r_-^{*}$, and the transition becomes second order. For $\alpha>2$ the transition for $\varphi^*$ is then always second order as $c^*_{\text{cr}} =0$. For $1<\alpha<2$ the transition for $\varphi^*$ can change from second to first order in the parameter space of $r_+^{(0)*},r_-^{(0)*}$ which determines $r_-^{*}$ in Eq.~\eqref{eq:SM:MF-reduced-2}, as $c^*_{\text{cr}}$ is determined by $\tilde{u}_1,\tilde{u}_2$ only.

To obtain the phase diagram of $\varphi^*$ in the parameter space $r_+^{(0)*},r_-^{(0)*}$, we focus on specific values of $\tilde{u}_1,\tilde{u}_2$ and solve numerically the MF equations~\eqref{eq:SM:MF-reduced} and \eqref{eq:SM:MF-reduced-2}. However, in the MF equation~\eqref{eq:SM:MF-reduced} a third quantity $\log(2\Lambda^2/\bar{v})$ enters. We shall set it to zero for now and show later that its value has no qualitative impact on the phase diagram. More concretely, we solve the MF equation~\eqref{eq:SM:MF-reduced} or \eqref{eq:SM:MF-reduced-3} with given $\tilde{u}_1,\tilde{u}_2$ for different values of $r_+^{(0)*}, r_-^{(0)*}$; if two solutions $c_1^*<c_2^*$ exist we always take $c_2^*$ as argued earlier, and if no solution exists we take $c^*=0$. We then substitute $c^*$ into Eq.~\eqref{eq:SM:MF-reduced-2} and obtain $r_-^{*}$ which is then used to compute $\varphi^*$ using Eq.~\eqref{eq:SM:order-parameter}. If $c^*<r_-^{*}$ we then take $\varphi^*=0$ meaning there is no vestigial AM order. For $\tilde{u}_1=1.6,\tilde{u}_2=1.9$, this gives Fig.~\ref*{M:fig:phasediagram} in the main text.

\subsection{The critical line of first order transitions}

Let us determine the critical line along which the transition is first order. As mentioned earlier, this is determined by the constant $\bar{r}_{0,\text{cr}}$ which is the maximum of the function $f(x)$ for given $\tilde{u}_1,\tilde{u}_2$. This gives a straight line in the parameter space of $r_+^{(0)*},r_-^{(0)*}$:
\begin{equation}\label{eq:SM:criticalline-1}
     r_+^{(0)*} -\frac{r_-^{(0)*}(\tilde{u}_1-\tilde{u}_2)}{2+\tilde{u}_1+\tilde{u}_2}=\bar{r}_{0,\text{cr}}.
\end{equation}
Remember we have set $\log(2\Lambda^2/\bar{v})=0$. However, $r_-^{*}$ increases with $r_-^{(0)*}$ and at sufficiently large $|r_-^{(0)*}|$ we shall always have $c^*_{\text{cr}} < r_-^{*}$ since $c^*_{\text{cr}}$ is independent of $r_+^{(0)*},r_-^{(0)*}$. Then according to Eq.~\eqref{eq:SM:order-parameter} and the preceding discussions, the $\varphi^*$ transition becomes second order: the critical line for the first order transition of $\varphi^*$ terminates far away from $|r_-^{(0)*}|$. This is demonstrated by the white dashed lines in Fig.~\ref*{M:fig:phasediagram}(b). Finally we note that a non-zero $\log(2\Lambda^2/\bar{v})$ will change the horizontal intercept of the critical line and the $\varphi^*$ values, but cannot change the qualitative behavior demonstrated above.

\subsection{Critical index}

We have demonstrated that a second order transition for $\varphi^*$ can only occur only when $c^*$ is past the transition point at certain intermediate values $c_0^* = r_{-,\text{cr}}^{*}$. We now show that the critical index for $\varphi^*\propto (T_c-T)^\alpha$ still has the MF value $1/2$.

Let us take temperature as the parameter and assume that we have reached  the critical temperature $T_c$ at which $c_0^* = r_-^{(0)*}$. $c^*$ and $\varphi^*$ increase upon a further infinitesimal decrease of $\bar{r}_0$ which can be written $\delta \bar{r}_0 \propto( T_c-T)$. Accordingly let us linearize the MF equation \eqref{eq:SM:MF-reduced-3} near $c_0^*$:
\begin{equation}
 f'(c^*_0) \delta c^* =  \delta \bar{r}_0\rightarrow \delta c^* \propto (T_c-T). 
\end{equation}
We then substitute this into Eq.~\eqref{eq:SM:order-parameter} expanded around $c^*_0=r_{-,\text{cr}}^{*}$:
\begin{equation}
\begin{split}
    \varphi^* \approx \sqrt{2c^*_0 (1- \p_{c^*} r_{-,\text{cr}}^{*}) \delta c^*} \propto (T_c-T)^{1/2}.
\end{split}
\end{equation}
This proves our statement. For the second order transition to occur, it is required that $1- \p_{c^*} r_{-,\text{cr}}^{*}>0$, i.e. the change in $c^*$ must be larger than the change in $r_-^{*}$ so that the term inside the square root in Eq.~\eqref{eq:SM:order-parameter} becomes positive.

\section{Spectral functions}

For the two-orbital model Eq.~(\ref*{M:eq:model-electron}), the one-loop self-energy diagram Fig.~\ref*{M:fig:spectralweight}(a) is given by:
\begin{equation}\label{eq:SM:propagator}
\begin{split}
    \Sigma(p_0,\mathbf{p}) = T\sum_{q_0}\int\frac{\diff^2q}{(2\pi)^2} G(q_0,\mathbf{q}+\mathbf{Q}) \{D_{MM}(p-q)+ D_{NN}(p-q)+ \tau^zs^z[D_{MN}(p-q)+ D_{NM}(p-q)]\},
\end{split}
\end{equation}
where $p-q = (p_0-q_0,\mathbf{p}-\mathbf{q})$ and $q_0$ is the fermionic Matsubara frequency. Note the $\mathbf{Q}=(\pi,\pi)$ momentum carried by the $M, N$ bosons. $G(q_0,\mathbf{q})$ is the free electron Green's function which is diagonal in both spin and orbital indices:
\begin{equation}
    G(q_0,\mathbf{q}) = -\int_0^{1/T} \langle \text{T}_\tau \{ \Psi_\mathbf{q} (\tau) \Psi_\mathbf{q}^\dagger(0) \} \rangle e^{iq_0\tau}\diff \tau = (iq_0 - H_0+\mu)^{-1},
\end{equation}
where $H_0$ is the free electron Hamiltonian (\ref*{M:eq:model-electron}). This allows $\tau^z, s^z$ vertices to be moved the left in Eq.~\eqref{eq:SM:propagator}. Using the Lagrangian (\ref*{M:eq:boson-Lagrangian}, the boson propagators are given as the following inverse matrix in frequency-momentum space:
\begin{equation}
    \begin{pmatrix}D_{MM}(i\omega_n,\mathbf{q}) & D_{MN}(i\omega_n,\mathbf{q})  \\ D_{NM}(i\omega_n,\mathbf{q}) & D_{NN}(i\omega_n,\mathbf{q})\end{pmatrix} = 
    \begin{pmatrix} \omega_n^2 + r_{10} +q^2 & -\varphi_0 \\ -\varphi_0 &\omega_n^2 + r_{20} +q^2\end{pmatrix}^{-1}.
\end{equation}
Performing the standard summation over $q_0$ and analytical continuation $ i p_0\rightarrow \omega + i\delta$ gives for the imaginary part of the electrons with spin index $s=\pm 1$ and orbital index $\lambda=\pm 1$:
\begin{equation}\label{eq:SM:imaginarySE}
    \Im \Sigma_{\lambda s}^R(\omega,\mathbf{p}) =  - \int \left(\tanh \frac{\varepsilon}{2T}- \coth \frac{\varepsilon-\omega}{2T}\right)\Im G_{\lambda s}^R(\varepsilon,\mathbf{q}+\mathbf{Q}) \Im D_{\lambda s}^R(\omega-\varepsilon,\mathbf{p}-\mathbf{q})  \frac{\diff \varepsilon}{2\pi} \frac{\diff^2q}{(2\pi)^2},
\end{equation}
where the retarded boson propagator:
\begin{equation}
   D_{\lambda s}^R(\varepsilon,\mathbf{q}) = D_{MM}^R(\varepsilon,\mathbf{q})+ D_{NN}^R(\varepsilon,\mathbf{q})+ \lambda s [D_{MN}^R(\varepsilon,\mathbf{q})+ D_{NM}^R(\varepsilon,\mathbf{q})].
\end{equation}
For the two-orbital system we have:
\begin{equation}
   \Im G_{\lambda s}^R(\varepsilon,\mathbf{q}+\mathbf{Q})  = -\pi \delta[\varepsilon-\varepsilon_{\lambda s}(\mathbf{q}+\mathbf{Q})+\mu ].
\end{equation}
Writing $\xi_{\lambda s,\mathbf{q}+\mathbf{Q}} = \varepsilon_{\lambda s}(\mathbf{q}+\mathbf{Q})-\mu$ and integrating over $\omega$ in Eq.~\eqref{eq:SM:imaginarySE} gives:
\begin{equation}
    \Im \Sigma_{\lambda s}^R(\omega,\mathbf{p}) =  \frac{1}{2} \int \left(\tanh \frac{\xi_{\lambda s,\mathbf{q}+\mathbf{Q}}}{2T}- \coth \frac{\xi_{\lambda s,\mathbf{q}+\mathbf{Q}}-\omega}{2T}\right) \Im D_{\lambda s}^R(\omega-\xi_{\lambda s,\mathbf{q}+\mathbf{Q}},\mathbf{p}-\mathbf{q}) \frac{\diff^2q}{(2\pi)^2}.
\end{equation}
Evaluating the momentum integral numerically at $\omega=0$ gives the result in the paper. For the positive infinitesimal $\delta$ we used $\delta = 0.001t$. Since $\tanh (\varepsilon/2T), \coth (\varepsilon/2T) \rightarrow \sign \varepsilon$ at $\varepsilon \gg T$, the dominant contribution of the integral comes from the momentum region $|\xi_{\lambda s,\mathbf{q}+\mathbf{Q}}| \lesssim T$ as mentioned in the main text.

\begin{figure}
    \centering
    \includegraphics[width=0.5\linewidth]{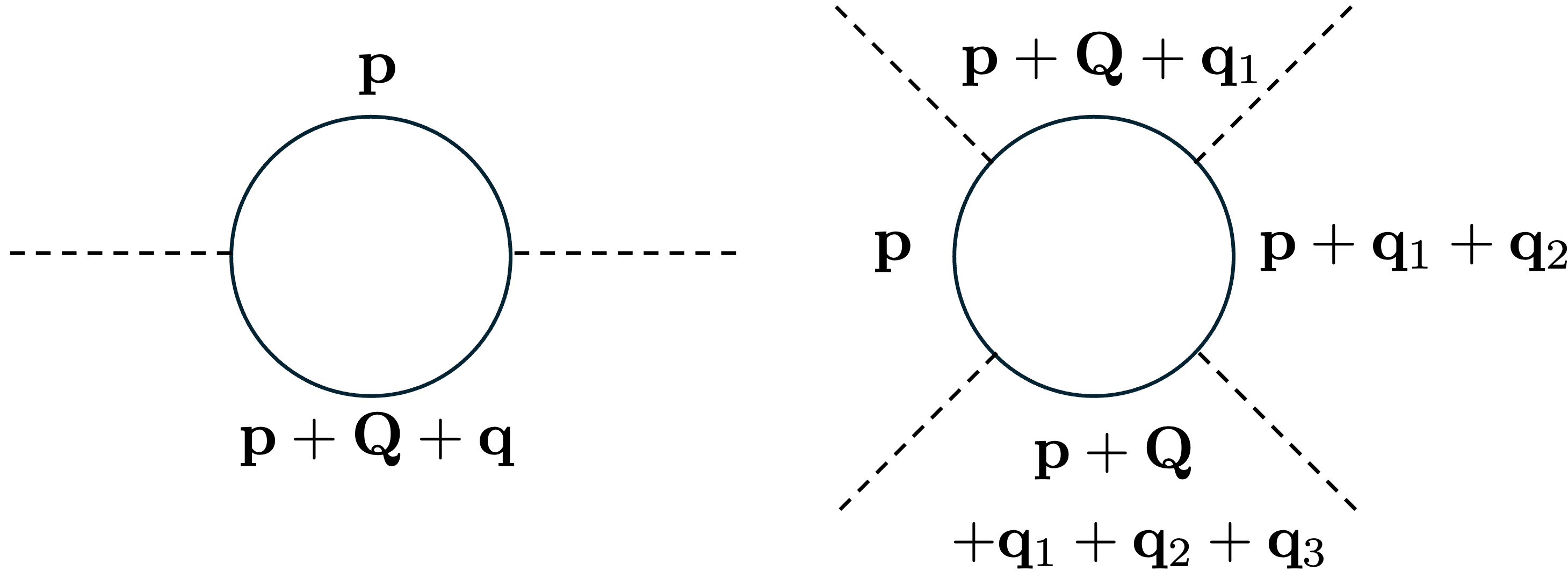}
    \caption{The diagrammatic expression for the first two terms in the free energy~\eqref{eq:SM:free-energy-2orbital}. The solid lines correspond to the free electron propagator $G_0$. The external dashed lines correspond to $V_{\text{el-b}}$ and carries additionally momentum $\mathbf{Q}$. }
    \label{fig:SM:free-energy-diagram}
\end{figure}

\section{Derivation of the free energy Eq.~(2) from microscopic Hamiltonians}

\subsection{The two-orbital model}

The model we consider is defined on a square lattice with the following free electronic Hamiltonian in orbital space~\cite{leeb2024spontaneous}:
\begin{equation}
\begin{split}
   &H_0(\mathbf{p}) = \begin{pmatrix} \varepsilon_x(\mathbf{p}) & \varepsilon_{xy}(\mathbf{p})   \\ \varepsilon_{xy}(\mathbf{p})& \varepsilon_y(\mathbf{p})\end{pmatrix} = \frac{\varepsilon_x+\varepsilon_y}{2}+ \tau^z\frac{\varepsilon_x-\varepsilon_y}{2}+ \tau^x \varepsilon_{xy}, \\
   &\varepsilon_x(\mathbf{p}) = -2 t_1 \cos p_x -2t_2 \cos p_y - 4t_3 \cos p_x \cos p_y, \\ 
   &\varepsilon_y(\mathbf{p})  =  -2 t_1 \cos p_y -2t_2 \cos p_x - 4t_3 \cos p_x \cos p_y, \\
   &\varepsilon_{xy}(\mathbf{p})  = -4t_4 \sin p_x \sin p_y.
\end{split}
\end{equation}
$\tau^a$ are Pauli matrices in orbital indices, $t_{1,2}$ are nearest-neighbour (N.N.) orbital overlaps and $t_{3,4}$ are next N.N. ones across the square diagonals. $t_4$ hoppings have opposite signs for the two diagonals. In Eq.~(\ref*{M:eq:model-electron}) in the main text, $t_1 =t-\delta t,\ t_2 = t+\delta t, \ t_3=t_4=0$. The interactions are given by:
\begin{equation}
    V = J_1\mathbf{S}_i.\mathbf{S}_j  + J_2 N_i^z N_j^z, \ M_i = \Psi_{\lambda s}^\dagger(\mathbf{r}_i) s^z_{s s'} \Psi_{\lambda s'}(\mathbf{r}_i)  \ N_i = \Psi_{\lambda s}^\dagger(\mathbf{r}_i) (\tau^z)_{\lambda\lambda'} \Psi_{\lambda' s}(\mathbf{r}_i).
\end{equation}
The summation is over N.N. sites. 

The free electron propagator:
\begin{equation}\label{eq:SM:GF-free}
    G_0 (i\omega,\mathbf{p}) = (i\omega- H_0)^{-1} =(g-\bm{\tau}.\mathbf{f})^{-1}= \frac{g+\bm{\tau}.\mathbf{f}}{g^2-f^2}, \ g= i\omega -  \frac{\varepsilon_x+\varepsilon_y}{2}, \ \mathbf{f} = \left[\varepsilon_{xy},0,\frac{\varepsilon_x-\varepsilon_y}{2}\right].
\end{equation}
We perform the standard MF decoupling in the spin and orbital channels:
\begin{equation}
    M_\mathbf{q} =J_1\sum_\mathbf{p} \langle \Psi_{\mathbf{p}+\mathbf{Q}+\mathbf{q}}^\dagger s^z \Psi_{\mathbf{p}} \rangle, \ N =J_2\sum_\mathbf{p} \langle \Psi_{\mathbf{p}+\mathbf{Q}+\mathbf{q}}^\dagger \tau^z \Psi_{\mathbf{p}}\rangle,
\end{equation}
The interaction term now becomes Eq.~(\ref*{M:eq:el-boson-vertex}) in the main text. The free-energy is obtained by integrating out the electrons:
\begin{equation}\label{eq:SM:free-energy-2orbital}
    F = -\Tr \log (1-G_0V_{\text{el-b}}) =   \sum_n\frac{1}{n}\Tr[(G_0V_{\text{el-b}})^n],
\end{equation}
which corresponds to the one-loop approximation in Fig.~\ref{fig:SM:free-energy-diagram}, where each vertex $V_{\text{el-b}}$ corresponds to an external line with external momentum $\mathbf{Q}$. The trace is taken over spin, orbital indices and there is also a summation over the loop frequency and momenta. Up to quartic order this gives the free energy (\ref*{M:eq:free-energy}). Let us first consider the second order term:
\begin{equation}
\begin{split}
    F_2&=  \frac{T}{2} \sum_{n, \mathbf{p}} \tr\{ (s^zM_{-\mathbf{q}}+\tau^zN_{-\mathbf{q}})G_0(i\omega_n,\mathbf{p}+\mathbf{Q}+\mathbf{q}) (s^zM_\mathbf{q}+\tau^zN_\mathbf{q})G_0(i\omega_n ,\mathbf{p})\} \\ 
    &=T \sum_{n,\mathbf{p}} \frac{g g'+\mathbf{f}.\mathbf{f}'}{(g^2-f^2)(g'^2+f'^2)}  |M_\mathbf{q}|^2 + T \sum_{n,\mathbf{p}} \frac{g g'-f_xf_x'+f_zf_z'}{(g^2-f^2)(g'^2-f'^2)}  |N_\mathbf{q}|^2, \ g'=g(\mathbf{p}+\mathbf{Q}+\mathbf{q}), \ f_i' = f_i(\mathbf{p}+\mathbf{Q}+\mathbf{q}).
\end{split}
\end{equation}
We see that odd powers of $V_{\text{el-b}}$ vanish due to momentum conservation. We see that generally in the even power terms, in taking the trace over spin indices, odd numbers of $s^z$ gives zero and only even powers of $M$ can appear. The $\mathbf{q}=0$ terms give the quadratic coefficients of the free energy, and expanding in small $\mathbf{q}$ gives the $q^2$ gradient terms. For the quartic power terms, it is convenient to define $\tilde{\mathbf{f}}=(-f_x,0,f_z)$. Rather long calculations give:
\begin{eqnarray}
    u_1 &=& \sum_{n,\mathbf{p}}\frac{2\left[ g^4 + g^2(f^2+f'^2+3 \mathbf{f}.\mathbf{f}') + f^2 f'^2\right] }{(g^2-f^2)^2(g^2-f'^2)^2},  \\
    u_2&=& \sum_{n,\mathbf{p}} \frac{2g^4 + 2g^2(4\tilde{\mathbf{f}}.\mathbf{f}' + f^2+\tilde{f}'^2) + 2[2(\tilde{\mathbf{f}}.\mathbf{f}')^2-f^2 \tilde{f}'^2]}{(g^2-f^2)^2(g^2-f'^2)^2}. 
\end{eqnarray}
For the term proportional to $M^2N^2$, from the four vertices of the form $(Ms^z +  N\tau^z)$ we need two $\tau^z$ to give $N^2$. The total number of such terms is $C^2_4 = 6$. Due to the cyclic property of trace we can visualise this as a necklace with four sockets to place two beads. Each interval between two sockets corresponds to a Green's function $(g+\bm{\tau}.\mathbf{f})$. There are two distinct configurations: $2$ where the beeds are opposite and $4$ where the beads are adjacent to each other. The first configuration gives:
\begin{equation}
     \frac{2}{(g^2-f^2)^2(g^2-f'^2)^2}\times  \tr [(g+\bm{\tau}.\mathbf{f})(g+\bm{\tau}.\mathbf{f}')\tau^z(g+\bm{\tau}.\mathbf{f})(g+\bm{\tau}.\mathbf{f}')\tau^z],
\end{equation}
For the second configuration:
\begin{equation}
     \frac{2}{(g^2-f^2)^2(g^2-f'^2)^2}\times \left\{ \tr [(g+\bm{\tau}.\mathbf{f})\tau^z (g+\bm{\tau}.\mathbf{f}')(g+\bm{\tau}.\mathbf{f})(g+\bm{\tau}.\mathbf{f}')\tau^z]  + \tr [(g+\bm{\tau}.\mathbf{f})(g+\bm{\tau}.\mathbf{f}')(g+\bm{\tau}.\mathbf{f})\tau^z (g+\bm{\tau}.\mathbf{f}')\tau^z] \right\}.
\end{equation}
These two terms can be evaluated by commuting $\tau^z \bm{\tau}.\mathbf{f} =\bm{\tau}.\tilde{\mathbf{f}} \tau^z $ and $(\tau^z)^2=1$, then using the identity $\tr (\tau^a\tau^b\tau^c\tau^d) = 2(\delta_{ab}\delta_{cd} -\delta_{ac}\delta_{bd}+\delta_{ad}\delta_{bc})$.

\subsection{three-orbital model}

\begin{figure}
    \centering
    \includegraphics[width=0.2\linewidth]{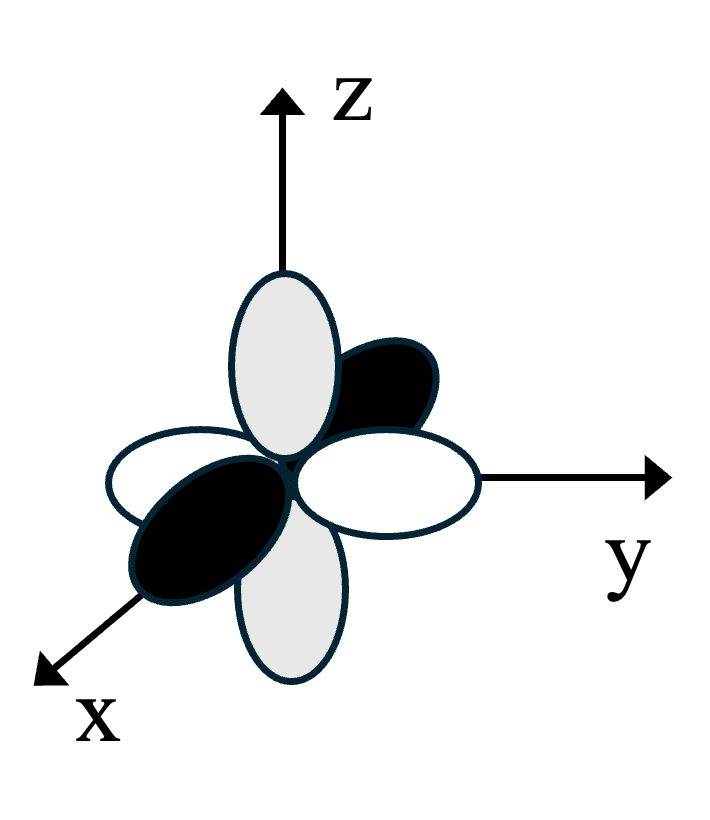}
    \caption{The graphic representation of the $3$ orbitals at each site. The $z$-orbital (in gray) is perpendicular to the $xy$-plane and is invariant under a C$_4$ rotation. }
    \label{fig:SM:zorbital}
\end{figure}

Let us now consider the three-orbital model in Ref.~\cite{kaushal2026spontaneous}. Compared to the two-orbital model, there is now the additional $\lambda = z$ orbital shown graphically in Fig.~\ref{fig:SM:zorbital}: $\Psi_{s\mathbf{p}} = (\Psi_{xs,\mathbf{p}},\Psi_{ys,\mathbf{p}},\Psi_{zs,\mathbf{p}})$ where $s$ is the spin index. The free Hamiltonian in orbital space is:
\begin{equation}
    H_0(\mathbf{p}) =\sum_{s,\mathbf{p}}\Psi_{s\mathbf{p}} ^\dagger \begin{pmatrix}\varepsilon_1(p_x) &0&0\\ 0& \varepsilon_1(p_y) & 0 \\ 0&0&\varepsilon_2(\mathbf{p}) +\Delta  \end{pmatrix}\Psi_{s\mathbf{p}};\ \varepsilon_1(p) = -2t \cos p, \ \varepsilon_2(\mathbf{p}) = -2t_z (\cos p_x+p_y).
\end{equation}
Here $t, t_z$ are the N.N. intra-orbital hopping terms for $x,y$ and $z$ orbitals respectively and inter-orbital transitions are neglected. The $\Delta$ term is the on-site potential for the $z$-orbitals.

The on-site interaction term in Ref.~\cite{kaushal2026spontaneous} is given by:
\begin{equation}
    V = U\sum _\alpha n_{\alpha, \uparrow} n_{\alpha,\downarrow} + U'\sum_{\alpha\ne \beta =x,y,z} n_\alpha n_\beta - J \sum_{\alpha\ne \beta}\mathbf{S}_{\alpha}.\mathbf{S}_\beta. 
\end{equation}
The first and the second terms are the intra- and inter-orbital Hubbard couplings. The third term is the intra-orbital ferromagnetic Hund's coupling between the orbitals. Since we are not interested in superconductivity, we ignore a pairing term that is present in Ref.~\cite{kaushal2026spontaneous}.

In the MF decoupling, we decouple the following channels:
\begin{equation}\label{eq:SM:MF-3-orbital}
    M = \sum_{\lambda = x,y} \langle \Psi_\lambda^\dagger s^z \Psi_\lambda \rangle, \ N =  \langle \Psi^\dagger \tau^z \Psi \rangle, \ m = \langle \Psi_z^\dagger s^z \Psi_z \rangle, \ \eta = \langle \Psi^\dagger \tau^z s^z \Psi \rangle = \langle \Psi_x^\dagger s^z \Psi_x \rangle-  \langle \Psi_y^\dagger s^z \Psi_y \rangle,\ \tau^z = \begin{pmatrix} 1 & 0 & 0\\ 0&-1& 0 \\ 0&0&0 \end{pmatrix}.
\end{equation}
Here $\tau^z$ acts on the orbitals $x,y$ only and $M$ is the spin density from the $x,y$ orbitals only. There is additionally the spin density from the $z$-orbital $m$ which helps to stabilize the AFM order via the ferromagnetic Hund's coupling. $\eta$ is the AM order parameter mentioned in the main text which shall be discussed in more detail in Sec.~\ref{sec:SM:AM-MF}

\subsubsection{MF Hamiltonian}

We now perform the decoupling according to \eqref{eq:SM:MF-3-orbital} to obtain the MF Hamiltonian. It turns out to be more convenient to use the operator formalism of MF decoupling:
\begin{equation}
    : O^\dagger O : \rightarrow \langle O^\dagger \rangle  :O :  + :O^\dagger : \langle O \rangle - \langle O^\dagger \rangle\langle O \rangle,
\end{equation}
where $O$ is the operator to be decoupled and $:O:$ denotes normal ordering. To begin we shall consider the occupation number averages:
\begin{align}\label{eq:MF-ansatz-3orbital}
    &\corr{n_{x\uparrow}} = \frac{1}{2}(M+N+\eta+n_0), \ \corr{n_{x\downarrow}} = \frac{1}{2}(-M+N-\eta+n_0), \\
    &\corr{n_{y\uparrow}} = \frac{1}{2}(M -N-\eta+n_0), \ \corr{n_{y\downarrow}} = \frac{1}{2}(-M -N+\eta+n_0), \\
    &\corr{n_{z\uparrow}} = \frac{n_{z0}}{2}+m, \  \corr{n_{z\downarrow}} = \frac{n_{z0}}{2}-m.
\end{align}
These are related to the MF parameters by ($s^z_i = \langle\Psi_i^\dagger s^z \Psi_i\rangle$ where $i = x,y,z$ are orbital indices):
\begin{align}\label{eq:SM:MF-eq-3-orbital-1}
    &s^z_x = \frac{1}{2}(\corr{n_{x\uparrow}} - \corr{n_{x\downarrow}}) = \frac{1}{2}(N+\eta), \ s^z_y =  \frac{1}{2}(N-\eta), \\
    & s^z_x+s^z_y = M, \ \frac{1}{2} \corr{\tau^z} =\frac{1}{2} \sum_\sigma ( \corr{n_{x\sigma}}- \corr{n_{y\sigma}}) = N, \\
    & \langle \tau^z s^z\rangle = s^z_x-s^z_y = \eta, \ \corr{s^z_z} = m,\\
    &\corr{n_x} = n_0+N, \ \corr{n_y}=n_0-N, \ \corr{n_z}=n_{z0}.
\end{align}
The density MF parameters $n_0,n_{z0}$ renormalize the chemical potential $\mu$ and the on-site potential $\Delta$ for the $z$-orbital and we shall ignore them below.

We decouple each term in $V$ in turn. Since the equation of motion is linear, we verify for each term the free energy minimisation gives the correct MF equation from above. First the Hubbard term:
\begin{align}
    &U n_{\alpha, \uparrow} n_{\alpha,\downarrow} \rightarrow   U(\corr{n_{\alpha, \uparrow}} n_{\alpha,\downarrow} + \corr{n_{\alpha, \downarrow}} n_{\alpha,\uparrow} ) -  U \corr{n_{\alpha, \uparrow}} \corr{n_{\alpha, \downarrow}} \\
    &\alpha=x: \ \frac{U}{2}[n_{x\uparrow} (-M+N-\eta)+ n_{x\downarrow} (M+N+\eta)] -\frac{U}{4}[N^2 - (M+\eta)^2] \\
    &\alpha=y: \ \frac{U}{2}[n_{y\uparrow} (-M-N+\eta)+ n_{y\downarrow} (M-N-\eta)] -\frac{U}{4}[N^2 - (M-\eta)^2] \\
    &\alpha = z: \ -2U s^z_z  + Um^2. 
\end{align}
The last term gives the correct MF equation for $m$. Adding the first and the second lines:
\begin{equation}
    -U [s^z_x (M+\eta) +s^z_y (M-\eta)] + \frac{U}{2} \tau^z N -\frac{U}{2}N^2 +\frac{U}{2}(M^2+\eta^2). 
\end{equation}
Differentiating with respect to $M,N,\eta$ gives the correct MF eqs in \eqref{eq:SM:MF-eq-3-orbital-1}.

The second term (where summation over $\sigma$ in $n_\alpha$ is implied):
\begin{equation}
\begin{split}
    &U'\sum_{\alpha<\beta} n_\alpha n_\beta \rightarrow  U'\left[n_x (\corr{n_y}+\corr{n_z}) + \corr{n_x}(n_y+n_z) + \corr{n_y}n_z + \corr{n_z}n_y -\corr{n_x }(\corr{n_y}+\corr{n_z})- \corr{n_y}\corr{n_z}\right] \\
    &=U'(-N \tau^z +N^2) \rightarrow N = \frac{1}{2}\corr{\tau^z}.
\end{split}
\end{equation}

The last Hund coupling term:
\begin{equation}
\begin{split}
    -2J \sum_{\alpha<\beta} \mathbf{S}_\alpha.\mathbf{S}_\beta \rightarrow -2J[&s^z_x( \corr{s^z_y}+\corr{s^z_z})+ \corr{s^z_x}(s^z_y+s^z_z) - \corr{s^z_x}( \corr{s^z_y}+\corr{s^z_z}) +\corr{s^z_z}s^z_y+ \corr{s^z_y}s^z_z   -\corr{s^z_y}\corr{s^z_z})] \\
    =&-J [s^z_x(M-\eta +2m) + (s^z_y+s^z_z)(M+\eta) + (N-\eta)s^z_z +2m s^z_y \\
    &-\frac{1}{2}(M^2-\eta^2) -m (M+\eta) - (M-\eta)m] \\
    =& -J[(M+2m) (s^z_x+s^z_y) -\eta (s^z_x-s^z_y)+ 2M s^z_z ] + \frac{J}{2}(M^2-\eta^2) + 2Jm M.
\end{split}
\end{equation}
Minimisation gives:
\begin{align}
    M: \ 2m+M = s^z_x+s^z_y +2s^z_z, \ \eta: \ \eta = s^z_x-s^z_y, \ m: \ M = s^z_x+s^z_y.  
\end{align}
This again agrees with \eqref{eq:SM:MF-eq-3-orbital-1}.

Putting all terms together we get for the MF single particle part:
\begin{equation}
\begin{split}
    H_0 =& -M[(U+J) (s^z_x+s^z_y) +2J s^z_z] +N\left[\left(\frac{U}{2}-U'\right) (n_x-n_y)\right]+\eta \left[(J-U)(s^z_x-s^z_y)\right] \\
    &-2[Us^z_z + J (s^z_x+s^z_y )]m +\left(U'-\frac{U}{2}\right)N^2 +\left(U+J\right)\frac{M^2}{2}+\left(U-J\right)\frac{\eta^2}{2} +2JmM +Um^2
\end{split}
\end{equation}
Here only $M$ couples to $m$ because they both do not contain $\tau^z$. $N$ and $\eta$ cannot couple because $\eta$ has spin quantum numbers. 
Explicitly in the basis of each orbital:
\begin{equation}
    \begin{split}
        &\Psi_x^\dagger \left\{ -[(U+J)M  +(U-J)\eta+2Jm]s^z- \left(U'-\frac{U}{2}\right) N \right\} \Psi_x \\
        +&\Psi_y^\dagger \left\{ -[(U+J)M  -(U-J)\eta+2Jm]s^z +\left(U'-\frac{U}{2}\right) N \right\} \Psi_y \\
        -&2\Psi_z^\dagger(JM +Um )s^z \Psi_z +\left(2U'-U\right)\frac{N^2}{2} +\left(U+J\right)\frac{M^2}{2}+\left(U-J\right)\frac{\eta^2}{2} +2JmM +Um^2.
    \end{split}
\end{equation}
We rescale the field $N \rightarrow N/(2U'-U), \ M \rightarrow M/(U+J) , \  \eta\rightarrow \eta/(U-J), \ m \rightarrow m /{U} $:
\begin{equation}\label{eq:SM:MF-scheme-1}
    \begin{split}
        &\Psi_x^\dagger \left\{ -[M  +\eta+2(J/U)m]s^z- \frac{1}{2}N \right\} \Psi_x +\Psi_y^\dagger \left\{ -[M  -\eta+2(J/U)m]s^z+ \frac{1}{2}N\right\} \Psi_y \\
        -&2\Psi_z^\dagger\left(\frac{J}{U+J}M +m \right)s^z \Psi_z +\frac{N^2}{2(2U'-U)} +\frac{M^2}{2(U+J)}+\frac{\eta^2}{2(U-J)} +\frac{2J}{U(U+J)}mM +\frac{m^2}{U}.
    \end{split}
\end{equation} 

\subsubsection{Free energy}
We now integrate out the electrons using Eq.~\eqref{eq:SM:free-energy-2orbital}. The calculation is simplified compared to the two-orbital case because the free electron Green's function $G_0$ is diagonal in orbital indices and spin degenerate.

For the $z$-orbitals to quadratic order:
\begin{equation}
    \Psi_z^\dagger [m +(J/U)M ]s^z\Psi_z \rightarrow G_{0z}(p)G_{0z}(p+Q) [m^2 + (J/U)^2M^2].
\end{equation}
Here and below for brevity we ignore the summation over loop frequency and momentum. We also write in the four-momentum notation $p = (ip_0,\mathbf{p})$ and $Q=(0,\mathbf{Q})$ where $p_0$ is the fermionic Matsubara frequency.

For the $x,y$-orbitals:
\begin{equation}
    -\Psi^\dagger \{ [M + \eta\tau^z +(J/U)m].s^z + \tau^zN/2\}\Psi, \ \Psi=(\Psi_x,\Psi_y).
\end{equation}
The free Green's functions:
\begin{equation}
    G_0(p) = \begin{pmatrix}G_{0x}(p)  & 0 \\ 0&  G_{0y}(p)  \end{pmatrix},
\end{equation}
are diagonal in orbital space. Therefore in integrating out the fermions, the order of the terms can be exchanged in orbital space and we can move all $G_0$ to one side and take the trace over spin first with $\tr(s^as^b) = \delta^{ab}/2$. For example for the quadratic order:
\begin{align}
    &\frac{1}{2} \tr \{G_0^2([M + \eta\tau^z +(J/U)m]s^z + \tau^zN/2 )^2\} = \frac{1}{4}\tr G_0^2 [(M + \eta\tau^z +(J/U)m)^2+N^2]  \\ 
    =&\frac{1}{4}(G_{0x}^2+G_{0y}^2) [M^2+ (J/U)^2m^2 + 2(J/U) mM + \eta^2 + N^2]+\frac{1}{2}(G_{0x}^2-G_{0y}^2)[(J/U) m\eta+  M\eta].
\end{align}
The second term vanishes after integrating over momentum due to $C_4$ symmetry.

The third order (the non-zero terms over spin trace are given by even powers of $s^z$):
\begin{align}
   &\frac{1}{3}\tr \{G_0^3([M + \eta\tau^z +(J/U)m]s^z + \tau^zN/2 )^3\} \\
   =&\frac{1}{6}\tr \{G_0^3 \tau^z N(N^2/4 + [M +\eta\tau^z +(J/U)m]^2 )\}.
\end{align}
Here again clearly the $\tau^z$ terms give zero upon trace. Therefore we have:
\begin{equation}
    \frac{1}{3}(G_{0x}^2(p)G_{0x}(p+Q)+ G_{0y}^2(p)G_{0y}(p+Q))[(J/U)m\eta N + M\eta N].
\end{equation}
The momentum argument is due to the momentum $\mathbf{Q}$ carried by $m, M, N$.

The above expressions can be anticipated on symmetry grounds. To leading order of $1/U$, the electron-boson coupling has the form:
\begin{equation}
    V_{\text{el-b}}'= c_1 \sum_{\lambda = x,y}  M\Psi_\lambda^\dagger s^z \Psi_\lambda + c_2 N \Psi^\dagger \tau^z \Psi + c_3  m \Psi_z^\dagger \tau^z \Psi_z,
\end{equation}
where $c_i$ are constants in terms of $U,U'$ and $J$. In integrating out the electrons using Eq.~\eqref{eq:SM:free-energy-2orbital}, we get the standard quartic terms for $M,N$ in the free energy (\ref*{M:eq:free-energy} in the main text. However, upon including $1/U$ terms, in the free energy $m$ can couple directly to $M$ as they both have spin quantum numbers. Since $N$ changes sign under C$_4$ whereas both $m$ and $M$ are invariant, the lowest-order coupling between $m$ and $N$ is given by $mM N^2$. Therefore to leading order in $m$, there appears additionally terms of the form:
\begin{equation}
   c_1' M m + c_2' N^2 m M,
\end{equation}
where $c_i'$ are new constants. 

We shall argue that these terms merely renormalize the free energy used in the main text. For example, $m$ can be integrated out using the formula:
\begin{equation}
    \langle \exp [ - m (c_1' M  + c_2' N^2 M)]\rangle_m = \exp\left[ \frac{\langle m^2\rangle}{2} (c_1' M  + c_2' N^2 M)^2\right].
\end{equation}
Thus we see that the effect of $m$ is to renormalize the coefficients $r^{(0)}_{\text{AFM}}$ and $v$ in the free energy~(\ref*{M:eq:free-energy}) which has the same form as before.

\subsection{The AM MF parameter}\label{sec:SM:AM-MF}

Let us now discuss the effect of $\eta$. Recall that $\eta$ corresponds to the MF channel: 
\begin{equation}
    \eta = \langle \Psi^\dagger \tau^z s^z \Psi \rangle = \langle \Psi_x^\dagger s^z \Psi_x \rangle-  \langle \Psi_y^\dagger s^z \Psi_y \rangle.
\end{equation}
This term with zero ordering momentum has the same symmetry as the vestigial AM order parameter $\varphi$ but is generated directly by electrons. From above, $\eta$ couples to the other MF parameters via the following terms to leading order:
\begin{equation}\label{eq:SM:coupling}
    \eta m N, \ \eta M N, \ \varphi \eta.
\end{equation}
This is also clear on symmetry grounds as $\eta$ has spin and orbital quantum numbers. Again we can integrate out $\eta$ then $m$ which simply renormalize the parameters in the free energy~(\ref*{M:eq:free-energy}). We note that if there is vestigial order, i.e. $\langle\varphi\rangle \ne 0$, then the last term in Eq.~\eqref{eq:SM:coupling} immediately induces a non-zero $\langle \eta \rangle$ as well and vice versa. Indeed, from symmetry considerations both $\varphi$ and $\eta$ can characterize the vestigial AM phase. However, since $\eta$ directly couples to the electron Hamiltonian, a non-zero $\langle \eta\rangle$ directly induces a spin-split Fermi surface. Note also that to one-loop, $\eta$ fluctuations cannot generate a $\tau^z s^z$ term in the electron self-energy which is responsible for the spin-orbital-dependent quasiparticle decay rate.

\nobalance 


%